# Sitnikov-type solution for the motion of infinitesimal mass in BiER4BP

*or*

*Sitnikov-type motions of test particle in BiER4BP*

**Sergey Ershkov***, Plekhanov Russian University of Economics,

36 Stremyanny Lane, Moscow, 117997, Russia,

Scopus number 60030998, e-mail: sergej-ershkov@yandex.ru,

**Alla Rachinskaya**, Odessa I. I. Mechnikov National University,

2 Dvoryanskaya St., Odessa, Ukraine, e-mail: rachinskaya@onu.edu.ua

**Abstract**

In this paper, we present a new ansatz for approximated solving equations of motion of the infinitesimal mass $m$ in case of *bi-elliptic* restricted problem of *four* bodies (BiER4BP) (where three primaries $M_1$, $M_2$, $M_3$ are rotating around their common centre of mass on *elliptic* orbits with hierarchical configuration $M_3 < M_2 \ll M_1$).

A new type of the solving procedure is implemented here to obtain the coordinates of the infinitesimal mass $m$. Meanwhile, the system of equations of motion has been successfully explored with respect to the existence of semi-analytical (approximated) way for presentation of the solution.

We obtain as follows: 1) the solution for coordinates $\{x, y\} = \{0, 0\}$ is approximately satisfied both the first and second equations of motion if we take into consideration assumption $\{M_3, M_2\} \ll M_1$, 2) the expression for coordinate $z(f)$ is given by the equation of 2-nd order, which describes *Sitnikov's-type* approximated solution. It means that test particle is moving along the $z$-axis, outward the common barycenter of the system (but perpendicular to the plane of the mutual rotation of all the primaries).

# 1. Introduction, BiER4BP (bi-elliptic restricted problem of four bodies).

In the restricted three-body problem (R3BP), the equations of motion describe the dynamics of an infinitesimal mass $m$ under the action of gravitational forces effected by two celestial bodies of giant masses $M_{Sun}$ and $m_{planet}$ ($m_{planet} < M_{Sun}$), which are rotating around their common centre of mass on Keplerian trajectories. The small mass $m$ is supposed to be moving (as first approximation) inside of *restricted* region of space near the planet of mass $m_{planet}$ Arnold 1978 (but outside the Roche-lobe's region Duboshin 1968 for this planet). It is worth noting that there is a large number of previous and recent fundamental works concerning analytical development with respect to the R3BP equations which should be mentioned accordingly Arnold 1978, Duboshin 1968, Ferrari & Lavagna 2018, Moulton 1900, Szebehely 1967. We should especially emphasize the theory of orbits, which was developed in profound work Szebehely 1967 for the case of the circular restricted problem of three bodies (CR3BP) (primaries are rotating around their common centre of mass on *circular* orbits) as well as the case of the elliptic restricted problem of three bodies Ferrari & Lavagna 2018 (ER3BP, primaries are rotating around barycenter on *elliptic* orbits).

As the next step of the developing the topic under investigation, we should also mention the case of the circle restricted problem of *four* bodies (CR4BP) Moulton 1900 (where particular solutions were obtained at the simplifying assumption concerning the circle orbits in case of CR4BP), and also the case of *bi-elliptic* restricted problem of *four* bodies (BiER4BP) Chakraborty & Narayan 2019, Liu & Gong 2018.

According to approach, suggested in Liu & Gong 2018, we consider in the current research the couple, consisting of two primaries of low masses $\{M_2, M_3\}$ which are rotating around their common centre of mass on *elliptic* orbits (with their common centre of mass ($M_2 + M_3$) which is simulteneously rotating on *elliptic* orbit around the central giant planet or the Sun $M_1$ on the same plane with the couple of two primaries $\{M_2, M_3\} < M_1$); additionally we mean that $M_3 < M_2$.

So, the motions of the primaries are coplanar (but those of infinitesimal mass is not). We assume also that distance $a_2$ between $\{M_2, M_3\}$ much less than distance $a_1$ between barycenter of two primaries $\{M_2, M_3\}$ and the central giant planet or the Sun $M_1$

$$a_2 \ll a_1 \qquad (*)$$

As for the natural restriction (of physical nature), we also mean that infinitesimal mass $m$ is supposed to be moving outside the *double* Roche's limit Duboshin 1968 for each of the primaries (as first approximation, not less than 7-10 $R_i$ where $R_i$ is the radius of the $i$-th primary with mass $M_i$, $i = 1, 2, 3$).

## 2. **Equations of motion of BiER4BP.**

A non-uniformly rotating coordinate system is defined to describe the motion of the fourth particle in the elliptic *planar* restricted four-body problem Liu & Gong 2018:

- the origin $O$ is the center of mass of all the primaries;
- $z$ axis is perpendicular to the plane of the mutual rotation of all the primaries;
- $x$ axis is pointing from barycenter of three primaries $\{M_1, M_2, M_3\}$ to barycenter of primaries $\{M_2, M_3\}$ (in Liu & Gong 2018 it was formulated as follows: "$x$ axis points from primary $M_1$ to barycenter of primaries $\{M_2, M_3\}$"; both assumptions coincide if $\{M_2, M_3\} \ll M_1$);
- and $y$ axis forms a right triad with $x$ and $z$ axes.

The dynamical equations of motion of the fourth particle in the non-uniformly rotating coordinate system can be written as in Moulton 1900, Liu & Gong 2018 (we should note that Eqn. (3) in Moulton 1900 was obtained without assumption of finite bodies or primaries moving in circles with respect one to another with constant uniform rotation; so, all the derivation and solving procedure up to the

Eqn. (3) and Eqn. (3) themselves coincide completely to those refererred to Eqn. (1) in work Liu & Gong 2018)

$$\ddot{X} - 2\dot{f}\dot{Y} - (\dot{f})^2 X - \ddot{f}Y = \frac{\partial U}{\partial X},$$

$$\ddot{Y} + 2\dot{f}\dot{X} - (\dot{f})^2 Y + \ddot{f}X = \frac{\partial U}{\partial Y}, \quad (1)$$

$$\ddot{Z} = \frac{\partial U}{\partial Z},$$

where, $U$ is the potential function and $f$ denotes the true anomaly tracing the orbit of the barycenter of primaries $\{M_2, M_3\}$ around the main barycenter of three primaries $\{M_1, M_2, M_3\}$; let us note that if we have chosen an approximation $\{M_2, M_3\} \ll M_1$, the barycenter of three primaries $\{M_1, M_2, M_3\}$ should be close to the center of mass of central giant body $M_1$.

It is interesting to note that in work Chakraborty & Narayan 2019 the dynamical equations of motion of the fourth particle (see Eqns. (7)-(8) in Chakraborty & Narayan 2019) were also considered in the non-uniformly rotating coordinate system, but centered at the barycenter of primaries $\{M_2, M_3\}$ and depending on two types of true anomalies refererred to different processes of tracing the orbits (one of the true anomalies is associated with tracing an elliptic motion of the barycenter of primaries $\{M_2, M_3\}$ around the main barycenter of three primaries $\{M_1, M_2, M_3\}$, another is referred to the rotating of primaries $M_2$, $M_3$ around their common center of mass $\{M_2, M_3\}$). Nevertheless, we will follow in the current research by the approach, suggested in work Liu & Gong 2018 (and applied here for investigations of the equations of motion of massless particle, moving under the action of forces considered in the *plane* variant of BiER4BP).

The expression Liu & Gong 2018 for $U$ is given by

$$U = G \cdot \sum_{i=1}^{3} \left( \frac{M_i}{R_i} \right), \quad (2)$$

$$R_i = \sqrt{(X - X_i)^2 + (Y - Y_i)^2 + (Z - Z_i)^2},$$

where, $G$ is the Gaussian constant of gravitation; $R_i$ are distances of the infinitesimal mass $m$ from each of the primaries $M_i$ ($i$ = 1, 2, 3), respectively ($\{X_i, Y_i, Z_i\}$ are the coordinates of primaries $M_i$).

Unlike the CR4BP Moulton 1900, the positions of the primaries are not fixed in the rotating frame (in case of the elliptic restricted problem) - as they move along elliptical orbits, their relative distance ρ is not constant during a time

$$\rho = \frac{a_1 \cdot (1 - e_1^2)}{1 + e_1 \cdot \cos f} \qquad (3)$$

where, $e_1$ is the eccentricity of *elliptic* orbit of the rotating barycenter of primaries $\{M_2, M_3\}$ around their common centre of mass with primary $M_1$. By setting the scaling of mass, distance and time in such a way as in Liu & Gong 2018 that

$$\begin{cases} [M] = M_1 + M_2 + M_3 = M \\ [L] = \rho \\ [T] = [\rho^3/(G \cdot M)]^{\frac{1}{2}} = \sqrt{1 + e_1 \cdot \cos f} / \dot f \end{cases} \qquad (4)$$

we introduce by the transformation of the previous, non-uniformly rotating coordinate system $\{X, Y, Z\}$ (used in (1)), the *pulsating* coordinate system $\{x, y, z\}$ with new scaling; so, we have with the help of (4)

$$\begin{cases} X = \rho \cdot x \\ Y = \rho \cdot y \\ Z = \rho \cdot z \end{cases} \qquad (5)$$

Then, equations (1) become

$$\ddot{x} - 2\dot{y} = \frac{1}{1+e_1 \cdot \cos f} \frac{\partial \Omega}{\partial x},$$

$$\ddot{y} + 2\dot{x} = \frac{1}{1+e_1 \cdot \cos f} \frac{\partial \Omega}{\partial y}, \qquad (6)$$

$$\ddot{z} = \frac{1}{1+e_1 \cdot \cos f} \frac{\partial \Omega}{\partial z} - z,$$

where dot indicates (d/d $f$) in (6), $\Omega$ is the scalar function

$$\Omega = \frac{1}{2}(x^2 + y^2 + z^2) + U_r$$

$$U_r = \frac{\mu_1}{r_1} + \frac{\mu_2}{r_2} + \frac{\mu_3}{r_3}, \qquad (7)$$

and $\mu_i = (M_i/M)$ ($i = 1, 2, 3$), $r_i$ is the dimensionless distance between the infinitesimal mass $m$ and $i$-th primary, respectively (in Liu & Gong 2018 $m_1$ should be associated with $M_3$ in the denotations here, *vice versa* $m_3$ instead of $M_1$).

It is important to clarify appropriately the expressions for $r_i$ in (7) which should be used in (6) for the further calculations (taking into account expressions (2) and transformation of coordinates (4)-(5)). According to Liu & Gong 2018, they are given by

$$r_i = \sqrt{(x - x_i)^2 + (y - y_i)^2 + (z - z_i)^2}$$

$$x_1 = -(\mu_3 + \mu_2), \qquad y_1 = 0, \qquad z_1 = 0,$$

$$x_2 = \mu_1 + \frac{\mu_3}{\mu_3 + \mu_2} r\cos\theta, \quad y_2 = \frac{\mu_3}{\mu_3 + \mu_2} r\sin\theta, \quad z_2 = 0, \qquad (8)$$

$$x_3 = \mu_1 - \frac{\mu_2}{\mu_3 + \mu_2} r\cos\theta, \quad y_3 = -\frac{\mu_2}{\mu_3 + \mu_2} r\sin\theta, \quad z_3 = 0,$$

where, $r$ denotes the distance between primaries $M_2$, $M_3$ whereas $\theta$ denotes the angle between radius-vector from barycenter of primaries $\{M_2, M_3\}$ to the primary $M_3$ and the $Ox$ axis. The expression for $r$ can be written as (taking into account (3)):

$$r = \left(\frac{1}{\rho}\right)\frac{a_2 \cdot (1-e_2^2)}{1+e_2 \cdot \cos f_2} \qquad (9)$$

where, $a_2$ is the semi-major axis of binary system $\{M_2, M_3\}$, $M_3 < M_2$; $e_2$ is the eccentricity of *elliptic* orbit of the rotating primary $M_3$ around the barycenter of primaries $\{M_2, M_3\}$. The parameter $\theta$ is determined by

$$\theta = \theta_0 + f_2 - f \quad \Rightarrow \quad f_2 = (\theta - \theta_0) + f \qquad (10)$$

where, $\theta_0$ is the initial value of $\theta$. Meanwhile, from (9)-(10) we obtain

$$r = \left(\frac{1+e_1 \cdot \cos f}{a_1 \cdot (1-e_1^2)}\right)\frac{a_2 \cdot (1-e_2^2)}{\left(1+e_2 \cdot \cos((\theta-\theta_0)+f)\right)} \qquad (11)$$

Let us note also that in the current research we neglect the effect of variable masses of the primaries Singh & Leke 2010. As for the domain where the aforesaid infinitesimal mass $m$ is supposed to be moving, let us consider the Cauchy problem in the whole space. Finally, it is worth noting that both spatial ER3BP and ER4BP are not conservative, and no complete integrals of motion are known Liu & Gong 2018, Ershkov *et al*. 2020a, Llibre & Conxita 1990.

By appropriately transforming the right parts with regard to partial derivatives with respect to the proper coordinates $\{x, y, z\}$ Ershkov *et al*. 2020b, system (6) can be represented as (taking into account (7)-(11))

$$\ddot{x} - 2\dot{y} = \frac{1}{1+e_1 \cdot \cos f}\left(x - \frac{\mu_1}{r_1^3}(x-x_1) - \frac{\mu_2}{r_2^3}(x-x_2) - \frac{\mu_3}{r_3^3}(x-x_3)\right),$$

$$\ddot{y} + 2\dot{x} = \frac{1}{1+e_1 \cdot \cos f}\left(y - \frac{\mu_1}{r_1^3}(y-y_1) - \frac{\mu_2}{r_2^3}(y-y_2) - \frac{\mu_3}{r_3^3}(y-y_3)\right), \quad (12)$$

$$\ddot{z} = \frac{1}{1+e_1 \cdot \cos f}\cdot\left[z - \frac{\mu_1}{r_1^3}(z-z_1) - \frac{\mu_2}{r_2^3}(z-z_2) - \frac{\mu_3}{r_3^3}(z-z_3)\right] - z,$$

where

$$r_1 = \sqrt{(x+\mu_3+\mu_2)^2 + y^2 + z^2},$$

$$r_2 = \sqrt{(x-\mu_1 - \frac{\mu_3}{\mu_3+\mu_2}r\cos\theta)^2 + (y - \frac{\mu_3}{\mu_3+\mu_2}r\sin\theta)^2 + z^2}, \quad (13)$$

$$r_3 = \sqrt{(x-\mu_1 + \frac{\mu_2}{\mu_3+\mu_2}r\cos\theta)^2 + (y + \frac{\mu_2}{\mu_3+\mu_2}r\sin\theta)^2 + z^2},$$

## 3. Sitnikov's-type approximated solution for the system of Eqns. (12).

Let us obtain the numerical, approximated solution of equations (12) (we consider approximation $r \to 0$ which stems from the assumption (*), and $\{\mu_2, \mu_3\} \to 0$)

$$\ddot{x} - 2\dot{y} = \frac{1}{1+e_1 \cdot \cos f}\left(x - \frac{\mu_1}{((x+(\mu_2+\mu_3))^2 + y^2 + z^2)^{\frac{3}{2}}}(x+\mu_3+\mu_2) - \frac{\mu_3+\mu_2}{((x-\mu_1)^2 + y^2 + z^2)^{\frac{3}{2}}}(x-\mu_1)\right),$$

(14)

$$\ddot{y} + 2\dot{x} = \frac{1}{1+e_1 \cdot \cos f}\left(y - \frac{\mu_1}{((x+(\mu_2+\mu_3))^2 + y^2 + z^2)^{\frac{3}{2}}}y - \frac{\mu_3+\mu_2}{((x-\mu_1)^2 + y^2 + z^2)^{\frac{3}{2}}}y\right),$$

where, solution $\{x, y\} = \{0, 0\}$ is approximately satisfied both the equations (14) above (if we take into consideration assumption $\{\mu_2, \mu_3\} \to 0$).

In this case, we obtain from third of equations (12)

$$\ddot{z} = \frac{1}{1+e_1 \cdot \cos f} \cdot \left[z - \frac{\mu_1}{z^2} - \frac{(\mu_2+\mu_3)z}{((\mu_1)^2 + z^2)^{\frac{3}{2}}}\right] - z, \Rightarrow$$

$$\ddot{z} \cong \frac{1}{1+e_1 \cdot \cos f} \cdot \left[\frac{z^3 - \mu_1}{z^2}\right] - z, \qquad (15)$$

i.e., the *appropriate* equation of 2-nd order, which describes the solution for coordinate $z(f)$ (*Sitnikov's-type* Duboshin 1968 approximated solution of equations (12)). It means that test particle is moving along the *z*-axis, outward the common barycenter of the system (but perpendicular to the plane of the mutual rotation of all the primaries).

Let us demonstrate the numerical application to the real astrophysical problem (here and below we will consider modelling of the triple system "Sun-Earth-Moon" $\{M_1, M_2, M_3\}$). We should note that we have used for calculating the data the Runge–Kutta fourth-order method with step 0.0001 starting from initial values;

besides, we have chosen for our calculations (for modelling the triple system "Sun-Earth-Moon" $\{M_1, M_2, M_3\}$) as follows:

$e_1 = 0.0167$, $\mu_2 \cong (1/332946)$, $\mu_3 \cong (1/328901) - \mu_2$, $\mu_1 = 1 - (\mu_2 + \mu_3)$.

The results of numerical calculations for coordinate $z(f)$ we schematically imagine at Figs.1-3. As for the initial data for Eqn. (15), we have chosen as pointed below:

1) $z_0 = 1$, $(\dot{z})_0 = 0.8$

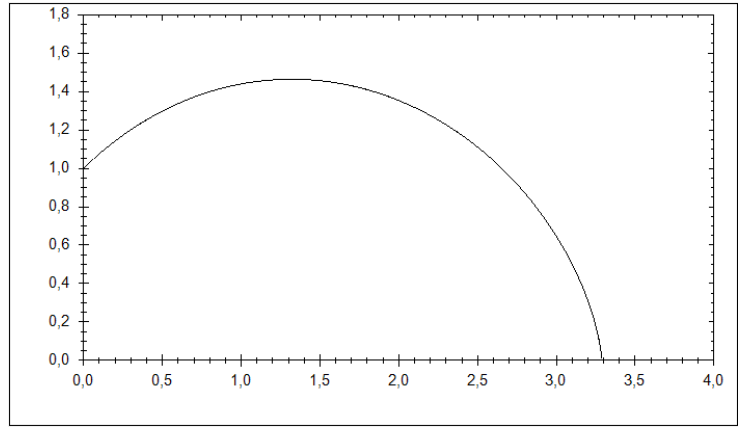

Fig.1. Results of numerical calculations of the coordinate $z(f)$

(as we can see from graphical plot of solution, $z(f) \to 0$ at $f \cong 3.3$ rad, which approximately corresponds to the angle $189.1° > \pi$).

2) $z_0 = 0.05$, $(\dot{z})_0 = 6.2$

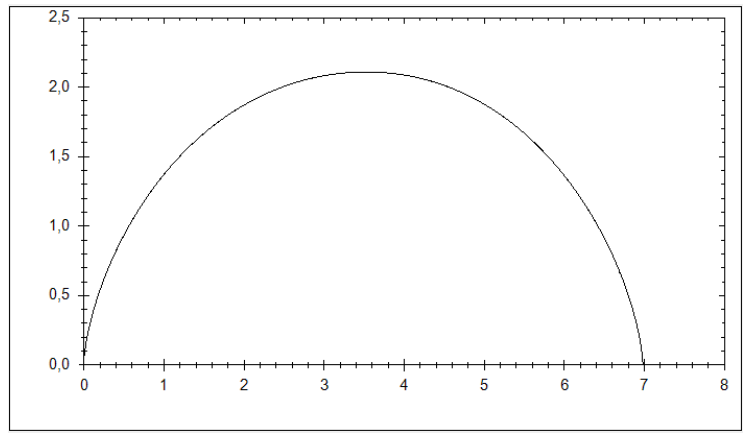

Fig.2. Results of numerical calculations of the coordinate $z(f)$

($z(f) \to 0$ at $f \cong 7$ rad, which approximately corresponds

to the angle $401° = 360+41°$).

3) $z_0 = 0.5$, $(\dot{z})_0 = 1.97$

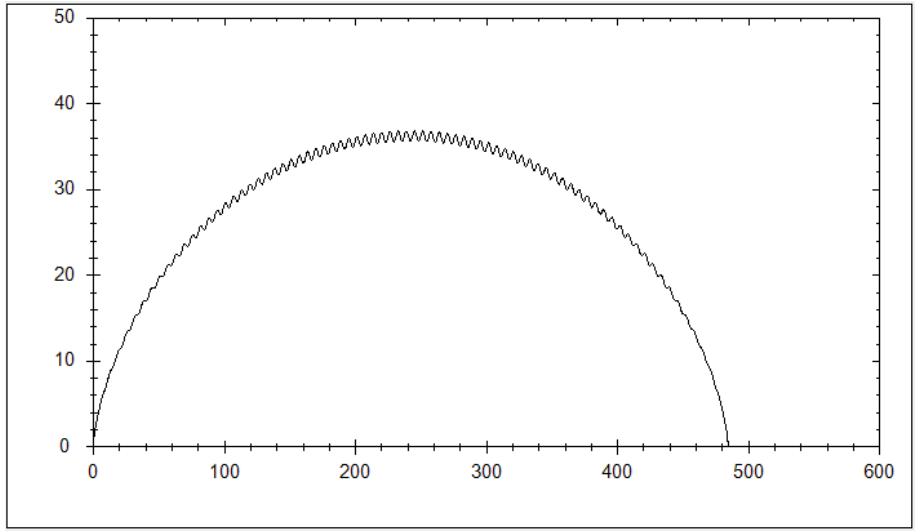

Fig.3. Results of numerical calculations of the coordinate $z(f)$

($z(f) \to 0$ at $f \cong 480$ rad, which approximately corresponds

to the angle $27502° = 76*(360°)+142°$.

## 4. Discussion.

As we can see from the derivation above, equations of motion (1) (in the form (12) Ershkov *et al*. 2020b) even for the appoximated problem (14) are proved to be very hard to solve analytically. Nevertheless, we have succeeded in obtaining from third equation of system (12) the equation (15) for the coordinate *z* (under mandatory condition that $\{x, y\} = \{0, 0\}$ is approximately satisfied both the first and second equations (14) if we take into consideration assumption $\{\mu_2, \mu_3\} \to 0$).

Thus, we have obtained *Sitnikov's-type* Duboshin 1968 approximated solution of equations (12). It means that test particle is moving along the *z*-axis, outward the common barycenter of the system (but perpendicular to the plane of the mutual rotation of all the primaries).

Ending discussion, let us note that coordinate *z* should be calculated under the *natural* restriction (of physical nature) so that the infinitesimal mass *m* is moving outside the *double* Roche's limit Duboshin 1968 for the Sun $M_1$ (not less than 10 $R_1$ where $R_1$ is the radius of Sun, $R_1 \cong 0.7$ million kilometers). If we obtain during the solving procedure the solutions closer to the Sun than the *double* Roche's limit, calculations should be stopped immediately at this forbidden zone. In this particlar case of modelling the triple system "Sun-Earth-Moon" (where 1 a.e. = 149.6 million kilometers), it means that there is restriction at choosing coordinate *z*: $|z| > 0.0468$.

As for the data of closest approaches of artificial satellite to the Sun during e.g. 'Parker Solar Probe'-mission in the years 2018-2025, see NASA 2018 (the closest approach during satellite's fly-by near the Sun should be 10 $R_1$ from the center of Sun, all the trajectories are supposed to be in the Ecliptic plane of Earth). So, motion not far from the barycenter in "planet-satellite-Sun" system Abouelmagd & Sharaf 2013, Singh & Umar 2014, Zotos 2015 (like the aforementioned Sitnikov's-type Duboshin 1968 approximated solution (15)) could be of practical interest in the sense of preventing collision trajectories with the surface of Sun in future missions, indeed.

## 5. Conclusion.

In this paper, we present a new ansatz for approximated solving equations of motion of the infinitesimal mass *m* in case of *bi-elliptic* restricted problem of *four* bodies (BiER4BP) (where three primaries $M_1$, $M_2$, $M_3$ are rotating around their common centre of mass on *elliptic* orbits with hierarchical configuration $M_3 < M_2 << M_1$). A new type of the solving procedure is implemented here to obtain the coordinates of the infinitesimal mass *m*. Meanwhile, the system of equations of motion has been successfully explored with respect to the existence of semi-analytical (approximated) way for presentation of the solution.

We obtain as follows: 1) the solution for coordinates $\{x, y\} = \{0, 0\}$ is approximately satisfied both the first and second equations of motion if we take into consideration assumption $\{M_3, M_2\} \ll M_1$, 2) the expression for coordinate $z(f)$ is given by the equation of 2-nd order, which describes *Sitnikov's-type* approximated solution. It means that test particle is moving along the *z*-axis, outward the common barycenter of the system (but perpendicular to the plane of the mutual rotation of all the primaries).

We have pointed out the optimizing procedure for the solutions $\vec{r} = \{x, y, z\}$ (the distance of the infinitesimal mass *m* from the primary $M_1$ should not be not less than 10 $R_1$ where $R_1$ is the radius of the primary with mass $M_1$ or the Sun).

The suggested approach can be used in future researches for optimizing the *Sitnikov's-type* maneuvers of spacecraft (in space) which is moving near the Sun $M_1$ belonging to the triplet of massive objects which are rotating around their common centre of mass on *elliptic* orbits with hierarchical configuration $M_3 < M_2 \ll M_1$.

The last but not least, we should especially note that in classical Sitnikov solution for the case of four-body problem Suraj & Hassan 2013, Zotos 2018, all the primaries should have the equal masses, but the aforementioned *Sitnikov's-type* solution (presented in the current research) is much more realistic for practical application in the real astophysical problems: the test particle of infinitesimal mass is moving along the *z*-axis, outward the common barycenter of the system (perpendicular to the plane of the mutual rotation of all the primaries), which are rotating around their common centre of mass on *elliptic* orbits with hierarchical configuration $M_3 < M_2 \ll M_1$.

Also, remarkable articles should be cited, which concern the problem under consideration Chernousko, *et al.* 2017, Ershkov & Shamin 2018, and Kushvah, *et al.* 2007. Finally, we should note that the aims, approach for obtaining the final results and results themselves of the current research differ from work Ershkov *et al.* 2020b (and so such the results should be considered as novel), despite the fact that general ansatz or the used mathematical (numerical) solving procedure is similar.

**Appendix, A (stability of the solutions $\{x, y\} \to \{0, 0\}$ of Eqns. (14)).**

Let us provide the checking of stability of the approximated solutions $\{x, y\} \to \{0, 0\}$ for system of Eqns. (14) (where we should take into consideration an additional simplifying assumption $\{\mu_2, \mu_3\} \to 0$); both equations of system (14) could be reduced in this case as below (where $z(f) \neq 0$ is the solution of Eqn. (15))

$$\ddot{x} - 2\dot{y} \cong \frac{1}{1+e_1 \cdot \cos f}\left(x - \frac{\mu_1}{z^3}(x+\mu_3+\mu_2) + \frac{(\mu_3+\mu_2)\mu_1}{\left((x-\mu_1)^2 + z^2\right)^{\frac{3}{2}}}\right),$$

$$\ddot{y} + 2\dot{x} \cong \frac{y}{1+e_1 \cdot \cos f}\left(1 - \frac{\mu_1}{z^3}\right),$$

(16)

and, finally, we obtain from (16) (assuming that $\{\mu_2, \mu_3\} \ll x \ll \mu_1$)

$$\ddot{x} - 2\dot{y} \cong \frac{x}{1+e_1 \cdot \cos f}\left(1 - \frac{\mu_1}{z^3}\right),$$

$$\ddot{y} + 2\dot{x} \cong \frac{y}{1+e_1 \cdot \cos f}\left(1 - \frac{\mu_1}{z^3}\right).$$

(17)

Let us consider stability of the solutions of the system above with zero right parts of both the equations (17) for the approximated solutions $\{x, y\} \to \{0, 0\}$

$$\begin{array}{ll} \ddot{x} - 2\dot{y} \cong 0, & \dot{x} - 2y \cong C_1, \\ & \Rightarrow \\ \ddot{y} + 2\dot{x} \cong 0, & \dot{y} + 2x \cong C_2, \end{array}$$

(18)

where, we should choose constants of integration $\{C_1, C_2\} = 0$ for the class of approximated solutions $\{x, y\} \to \{0, 0\}$. So, we obtain from system of equations (18) (we should choose $\{A, B\} = const \ll 1$ in the expressions (19) below):

$$\begin{cases} y \cong \dot{x}/2, \\ \ddot{x} + 4x \cong 0, \end{cases} \Rightarrow \{x = A\cos(2f) + B\sin(2f), \quad y = -A\sin(2f) + B\cos(2f)\} \quad (19)$$

Thus, approximated estimations at analysis of the dynamics of non-stationary motions for infinitesimal mass *m* in the vicinity of the approximated stationary solution $\{x, y\} \to \{0, 0\}$ reveals a stable circular 2D motion around this point in (*x*, *y*)-plane with the apropriate component of motion in *z*-coordinate of *Sitnikov-type* (described by Eqn. (15)).

The last but not least, we should mention that the effect of radiation pressure (from the Sun) due to the sun light in addition to mutual gravitational attraction must be considered in a future research regarding stability of motion of the infinitesimal mass *m* for the *Sitnikov-type's* approximated solutions of the *bi-elliptic* restricted problem of *four* bodies (BiER4BP).

## **Conflict of interest**

On behalf of all authors, the corresponding author states that there is no conflict of interest.

Remark regarding contributions of authors as below:

In this research, Dr. Sergey Ershkov is responsible for the general ansatz and the solving procedure, simple algebra manipulations, calculations, results of the article in Sections 1-4 and also is responsible for the search of approximated solutions.

Dr. Alla Rachinskaya is responsible for approximated solving the *non-linear*

ordinary differential equation (15) by means of advanced numerical methods as well as is responsible for numerical data of calculations and graphical plots of numerical solutions.

Both authors agreed with results and conclusions of each other in Sections 1-5.